\documentclass[twocolumn,showpacs,preprintnumbers,amsmath,amssymb]{revtex4}

\usepackage{graphicx}
\usepackage{dcolumn}
\usepackage{bm}

\begin{document}

\draft

\title{Spin-dependent transport in molecular tunnel junctions}
\author{J.~R.~Petta}
   \altaffiliation[]{Present address: Physics Department, Harvard
University, Cambridge, MA 02138}
\author{S.~K.~Slater}
\author{D.~C.~Ralph}
\affiliation{Laboratory of Atomic and Solid State Physics, Cornell
University, Ithaca, New York 14853}

\date{\today}

\begin{abstract}
We present measurements of magnetic tunnel junctions made using a
self-assembled-monolayer molecular barrier. Ni/octanethiol/Ni
samples were fabricated in a nanopore geometry. The devices
exhibit significant changes in resistance as the angle between the
magnetic moments in the two electrodes is varied, demonstrating
that low-energy electrons can traverse the molecular barrier while
maintaining spin coherence.  An analysis of the voltage and
temperature dependence of the data suggests that the spin-coherent
transport signals can be degraded by localized states in the
molecular barriers.

\end{abstract}


\maketitle

The field of molecular electronics has made a number of recent
advances, with measurements performed on single molecules, and
demonstrations that molecules can exhibit diode and transistor
behaviors
\cite{Reed,Kergueris,Weber,Metzger,Zhou_APL_97,Park,Liang,Nitzan}.
However, to date nearly all molecular-electronics experiments have
focused on charge transport, without taking advantage of the
electron's spin.  Exceptions are experiments on carbon nanotubes
contacted by ferromagnetic electrodes \cite{Tsukagoshi_Nature_99}
and spin transport of photo-excited ($>$ 1 eV) carriers through
organic linkers between semiconductor quantum dots
\cite{Awschalom_Science_03}. Molecular devices may be well-suited
for applications requiring spin manipulation because the relative
weakness of spin-orbit and hyperfine interactions in many
molecules, compared to conventional semiconductor systems, may
help to isolate the spin from external degrees of freedom. Here we
study the spin-polarized transport of electrons tunneling through
a barrier consisting of a self-assembled organic monolayer between
two magnetic electrodes. We find that spin coherence can be
maintained during the tunneling process, as demonstrated by
changes in resistance as we vary the relative orientations of the
magnetic moments in the two electrodes.  Our measurements
demonstrate that spin-coherent transport in molecular devices is
possible for low-energy electrons, as distinct from previous
demonstrations for photo-excited carriers
\cite{Awschalom_Science_03}. However, our molecular barriers are
not ideal. Strong voltage and temperature dependence of the
junction magnetoresistance (JMR) and time-dependent
telegraph-noise signals suggest that the device properties can be
affected by localized states in the molecular barriers.

We fabricate nanometer-scale tunnel junctions with self-assembled
molecular barriers using the nanopore technique
\cite{Ralls_APL_89}, employed previously to study conduction
through molecular layers in nonmagnetic tunnel junctions
\cite{Zhou_APL_97,Petta_APL_00,Wang}. First, we use electron-beam
lithography and a timed reactive-ion etch to fabricate a
bowl-shaped hole through a suspended silicon-nitride membrane. The
area of the holes is characterized by evaporating Cu onto both
sides of test samples without breaking vacuum, measuring the
resistance of the metal contacts formed through individual holes,
and estimating the area of the contact using the Sharvin formula,
$R=(h/e^{2})(2\pi/(k_{F}^{2}A))$ \cite{Sharvin}.   We find hole
diameters in the range of 5--10 nm. Once a set of holes is made,
the procedure to create the molecular tunnel junctions begins by
evaporating a 10 \AA \space Ti adhesion layer (which does not fill
the hole), followed by 1000 \AA \space of Ni and 500 \AA \space of
Cu onto the bowl-shaped side of the sample (the bottom side of the
schematic in the inset of Fig.\ 1(c)). After breaking vacuum, we
immerse the samples immediately in a 1 mM solution of octanethiol
in ethanol, transfer them to an Ar glove box, and leave them to
react for at least 48 hours to form a self-assembled monolayer
(SAM) of octanethiol on the Ni \cite{Mekhalif_Langmuir_03}. We
choose to use octanethiol because the transport properties of this
molecule in contact with non-magnetic electrodes have been studied
extensively \cite{Bumm,Wold,Cui,Wang}.  Before evaporation of the
top contact, we rinse the samples with ethanol and blow dry with
dry N$_{2}$. The samples are then transferred to an evaporator and
pumped to high vacuum (10$^{-7}$ Torr).  For most of the devices
described below we deposit 10 \AA \space of Ti, followed by 300
\AA \space of Ni or Co and a 1000 \AA \space Cu capping layer.
Because the Ti adhesion layer could damage the organic monolayer,
we have also fabricated samples without any Ti in the top contact
near the sample region.  Devices fabricated with and without Ti
had similar magnetoresistance curves. All top-contact evaporations
are performed at a low rate ($\sim$0.1--0.2 \AA/s) and with the
target cooled to 77 K to minimize damage to the molecular layer.
By convention, positive voltages ($V$) correspond to electron flow
from bottom to top in the device schematic in Fig.\ 1(c). Unless
specified otherwise, all transport measurements were taken at a
temperature of 4.2 K.

To characterize our monolayers, as well as to distinguish effects
due merely to the solvent or the processing procedures, rather
than the SAM, we fabricated control samples by immersing the
devices in pure ethanol solvent for 48 hours instead of
octanethiol solution. For both the control samples and the SAM
devices, we found a distribution of resistance values (Fig.\ 1).
All but one of 9 controls without octanethiols had resistances
below $h/e^2 =$ 25.8 k$\Omega$, indicating the presence of a
metallic contact \cite{Warning}.  Several of the SAM samples also
had low resistances, from which we infer the presence of metallic
shorts through the molecular layer. However, 23 of 29 octanethiol
devices had resistances greater than $h/e^2$, indicating electron
transport via a tunneling mechanism. The resistance values
clustered in the M$\Omega$ range are similar to previous
measurements of alkanethiol molecules using Au electrodes in a
nanopore geometry \cite{Wang}. In the following, we will focus
solely on the junctions with resistance greater than $h/e^2$.

\begin{figure}[t]
\begin{center}
\includegraphics[width=8.5cm]{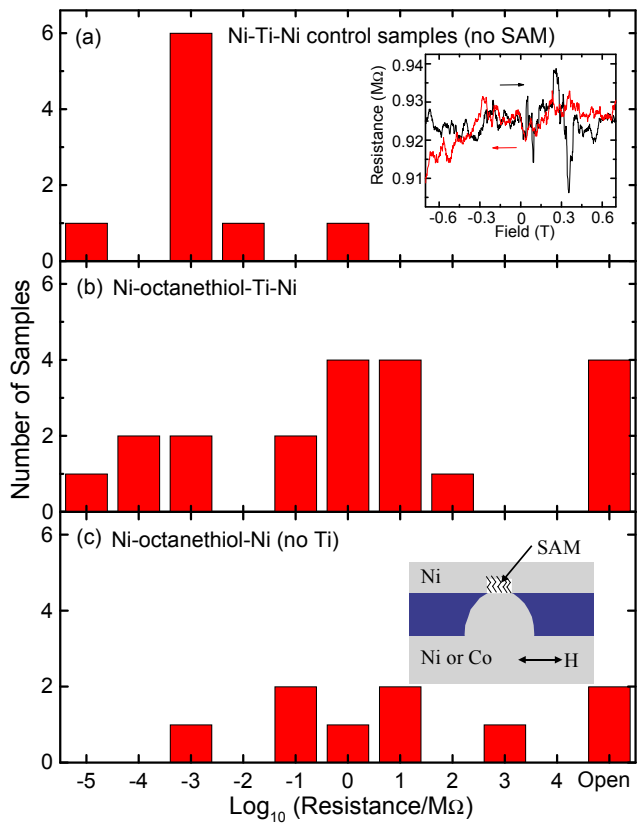}
\end{center}
\vspace{-0.6cm} \caption{Resistance histograms for different types
of samples at T=4.2 K and H=0 T. (a) Ni-Ti-Ni control samples in
which the samples were immersed in ethanol solvent for 48 h before
deposition of the Ti/Ni top contact. Inset: Magnetoresistance  of
the most resistive of the nine measured controls. The black curve
represents the increasing-field sweep and the red curve the
decreasing-field sweep. (b) Ni-octanethiol-Ti-Ni samples, (c)
Ni-octanethiol-Ni samples with no Ti adhesion layer in the sample
region. Inset: device geometry.} \vspace{-0.25cm}
\end{figure}

\begin{figure}[b]
\begin{center}
\includegraphics[width=8.7cm]{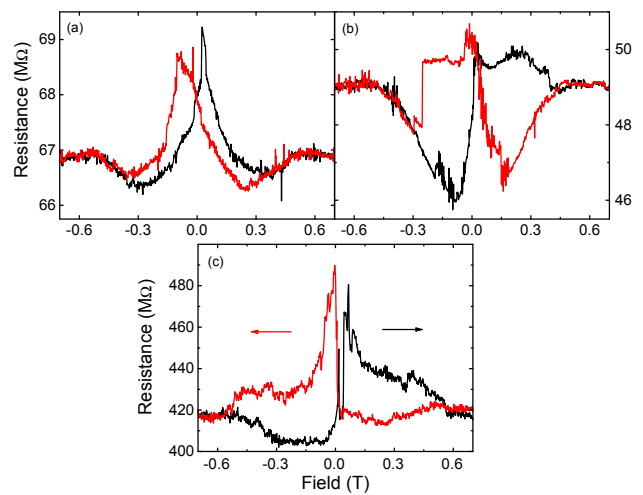}
\end{center}
\vspace{-0.6cm} \caption{R vs.\ H for three Ni-octanethiol-Ti-Ni
samples, all taken at 4.2 K. Black curves represent
increasing-field sweeps and red curves decreasing-field sweeps.
(a) Sample \#1, biased at 10 mV, (b) Sample \#2, biased at 10 mV,
(c) Sample \#3, biased at 5 mV.}\vspace{-0.5cm}
\end{figure}

In Fig.\ 2(a--c) we plot the resistance ($R=V/I$), measured at 4.2
K, as a function of magnetic field ($H$) applied in the sample
plane for three Ni-octanethiol-Ti-Ni samples with varying
resistances. For $|H|$$>$0.6 T, the magnetizations of the two
electrodes are parallel, resulting in an $H$-independent
resistance $R_P$.  We note that the saturation field is larger
than for simple planar Ni films, presumably due to the bowl-shaped
contact on one side of the device (inset, Fig.\ 1(c)). As $H$ is
swept through zero, the magnetizations of the two electrodes
undergo reversal processes at different fields, so that they may
approach an approximately antiparallel configuration before
ultimately aligning with the reversed field. During this process,
the samples exhibit clear changes in resistance, reaching values
of $R$ that are generally higher than $R_P$, although in some
samples for a range of $H$ the resistance may also dip below $R_P$
(e.g., Fig.\ 2(a--b)). If $R_{max}$ is the maximum resistance
measured and $R_{min}$ is the minimum, we define the positive and
negative junction magnetoresistances
$JMR_{+}$$=$($R_{max}$-$R_P$)/$R_P$ and
$JMR_{-}$$=$($R_{min}$-$R_P$)/$R_P$. For the three samples in
Fig.\ 2 we find $JMR_{+}$$=$3.5\%, $JMR_{-}$$=$-6.1\%, and
$JMR_{+}$$=$16.0\% for (a--c), respectively (see figure caption
for bias voltages). In comparison, the low resistance
Ni-ethanol-Ti-Ni control samples with no octanethiol barriers
exhibited much smaller resistance changes and qualitatively
different magnetoresistance traces (see the inset of Fig.\ 1(a)).

An estimate for the value of JMR corresponding to the mechanism of
direct electron tunneling through a barrier is given by the
Julliere formula, JMR = 2$P_1$$P_2$/(1-$P_1$$P_2$), where $P_1$
and $P_2$ are the tunneling spin polarizations associated with the
two electrodes \cite{Julliere_PLA_75}. For Ni, where
$P$$\approx$0.31, the Julliere estimate is JMR = 21\%
\cite{Monsma}. The largest JMR that we have measured is 16\%, for
a bias voltage of 5 mV (Fig.\ 2(c)).  This is 3/4 of the Julliere
value, from which we conclude that the electron spin is capable of
maintaining a high degree of coherence during the tunneling
process.

The measured values of JMR are strongly correlated with the sample
resistance, with the largest magnitudes associated with the most
resistive samples (Fig.\ 3(a)).  This suggests that imperfections
in the SAM, which should lower $R$, can also reduce the JMR. The
smaller JMR values  are likely due to contributions from transport
mechanisms that differ from simple direct tunneling through the
molecular barrier, as discussed below.

The quality of the SAM tunnel barriers can be characterized in
more detail by measurements of the voltage $(V)$ and temperature
$(T)$ dependence of their transport properties.  In Fig.\ 3(b), we
plot a high bias $I$-$V$ curve from a high-resistance
Ni-octanethiol-Ti-Ni junction as a function of $V$, along with a
fit to the Simmons tunneling model \cite{Wang,Simmons}. The fit
suggests a barrier height of 1.51$\pm$0.02 eV, in rough agreement
with the value of 1.4 eV measured for alkanethiols on gold
\cite{Wang}. We illustrate the $V$ and $T$ dependence of the JMR
in figures 4 and 5. The magnitude of the JMR depends strongly on
$V$ in all samples, and for some devices the JMR can change sign
to produce negative values in particular ranges of $V$ and $H$. In
all samples, the magnitude of the JMR decreases for large enough
values of $|V|$, and is typically reduced to less than 2\% for
$|V|$$>$40 mV (Fig.\ 4). The magnetoresistance is strongly
temperature dependent, as well (Fig.\ 4(c)).

Previously, qualitatively similar $V$ and $T$ dependences have
been observed for oxide tunnel barriers, for which the
characteristic voltage scale of the decaying magnetoresistance can
vary from 3 mV to 500 mV, depending on the barrier quality
\cite{Tsymbal_review}.  For oxide barriers, the mechanisms behind
negative JMR values and the $V$ and $T$ dependence of
magnetoresistance have been controversial, but the recent
observation that vacuum tunnel barriers give magnetoresistances
with very little $V$ dependence \cite{Wulfhekel} provides evidence
that two-step tunneling through localized states in the tunnel
barrier is a possible explanation for all three effects
\cite{Zhang_JAP,Tsymbal_PRL,Tsymbal_review}. We suggest that, in a
similar way, localized states in the octanethiol barrier may
explain many of the anomalous features that we measure as a
function of $V$ and $T$.

\begin{figure}[t]
\begin{center}
\includegraphics[width=8.5cm]{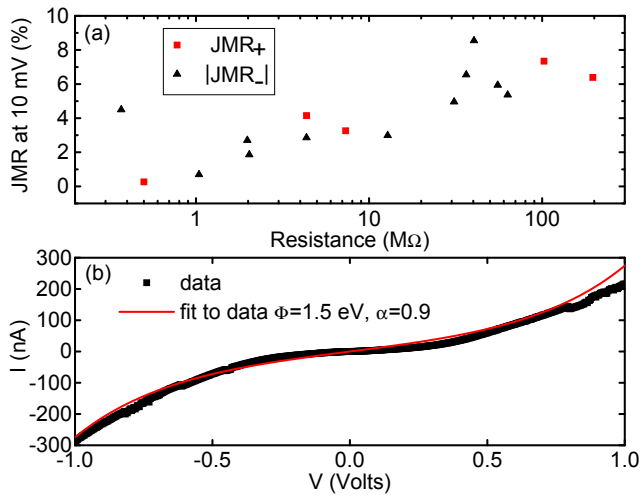}
\end{center}
\vspace{-0.8cm} \caption{(a) The largest of either $JMR_+$ or
$|$JMR$_-$$|$ for each sample at 10 mV bias, $T$=4.2 K plotted
versus the low-bias sample resistance. (b) $I$ versus $V$ for a
high resistance Ni-octanethiol-Ti-Ni junction at $T$=4.2 K and
$H$=0 T, along with a fit to the Simmons model with
$\Phi=1.51\pm0.02$ eV and $\alpha=0.90\pm0.01$. $\alpha$ takes
into account any asymmetry in the barrier profile (see ref. 12).}
\vspace{-0.5cm}
\end{figure}

Several of our samples exhibit striking time-dependent two-level
resistance fluctuations, {\it i.e.\ }telegraph noise (Fig.\ 5).
The frequency of the fluctuations varies as a function of both $V$
and $H$ \cite{Xiao}. The fluctuations could be due either to the
motion of electrons within localized charge defects or to changes
in the structural properties of the junctions.  In either case,
they demonstrate the importance of imperfections in the molecular
barrier.  The size of the changes in resistance suggest that a
very small number of molecules may be involved in these
fluctuations. For example, in sample \#2, the resistance at $H$=0
fluctuates by $\sim$0.5 M$\Omega$ out of a total resistance of 42
M$\Omega$ at $V$=-40 mV (see Fig.\ 5), for a fractional
fluctuation of $\sim$0.01. We estimate that the device area is in
the range 20--80 nm$^2$, based on the Sharvin resistance of test
nanopores filled with Cu. Assuming a packing density of $5$
molecules/nm$^{2}$ \cite{Widrig}, the device should therefore
contain 100--400 molecules. A fractional change of 0.01 therefore
corresponds to having 1--4 molecules switch their conductance
between fully off and on states, using the very rough assumption
that all molecules contribute equally to transport.  In other
devices, the amplitude of the telegraph signals corresponds to
fluctuations by a conductance equivalent to 1--12 molecules. These
numbers should be considered upper bounds, because if transport
through the barriers is non-uniform then the relative strength of
signals from individual molecules could be enhanced.

\begin{figure}[t]
\begin{center}
\includegraphics[width=8.5cm]{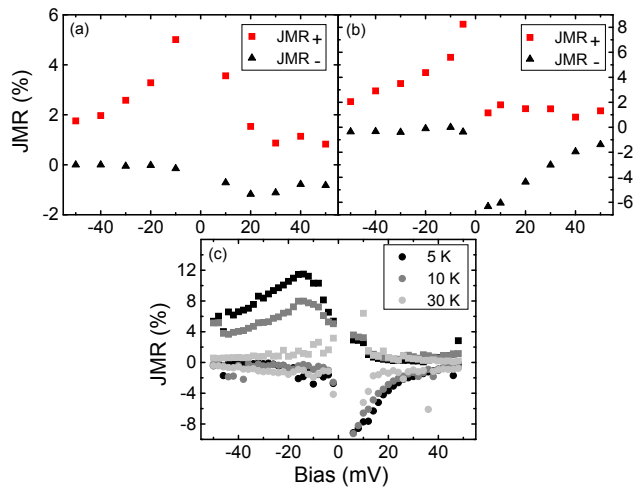}
\end{center}
\vspace{-0.8cm} \caption{$JMR_{+}$ and $JMR_{-}$ vs. bias for (a)
Sample \#1, (b) Sample \#2. For (a)--(b), T=4.2 K. (c) JMR$_+$
(squares) and JMR$_-$ (circles) plotted vs.\ bias and temperature
for Sample \#4. The sample shown in (c) has a low-bias resistance
of 36 M$\Omega$.} \vspace{-0.5cm}
\end{figure}

\begin{figure}[h]
\begin{center}
\includegraphics[width=8.5cm]{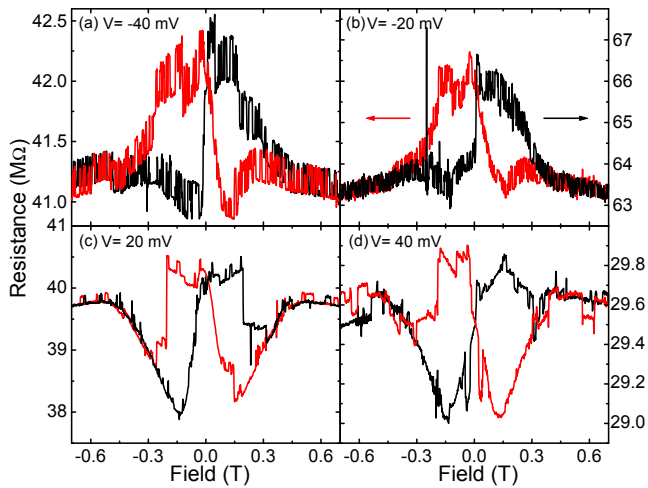}
\end{center}
\vspace{-0.5cm} \caption{R vs.\ H for Sample \#2 at four values of
bias for T=4.2 K.} \vspace{-0.5cm}
\end{figure}

In conclusion, we have fabricated and measured the transport
properties of magnetic tunnel junctions containing octanethiol
tunnel barriers. We find, first, that spin coherence can be
maintained in transport through these molecular devices. The
tunnel junctions exhibit JMR values of up to 16\% at low bias
voltages. This suggests that molecular bridges may prove useful in
applications involving electron-spin manipulation. However, we
also find a correlation between the JMR and the sample
resistances, strong $V$ and $T$ dependence for the
magnetoresistance, negative JMR values for particular ranges of
$V$ and $H$ in some samples, and the presence of telegraph noise.
All of these factors suggest the presence of localized states
within the octanethiol barrier
\cite{Zhang_JAP,Tsymbal_PRL,Tsymbal_review}. The origin of these
states is uncertain. They could be formed in the physical process
of top-layer deposition, by chemical reactions between the
monolayer and the metals, or by stress \cite{Son}. A better
understanding of the growth of SAMs on magnetic surfaces and
improved procedures for depositing top contacts are likely to
improve device yield and increase the JMR.

\begin{acknowledgments}
We thank R. A. Buhrman for discussions. This work was supported by
DARPA (N00173-03-1-G011), the ARO (DAAD19-01-1-0541), and the NSF
(DMR-0244713). A portion of the work was performed at the
NSF-funded Cornell Nanoscale Facility/NNIN.
\end{acknowledgments}


\begin{thebibliography}{99}
\bibitem{Warning} It is important to use ethanol as the solvent.
Other control samples made using THF as the solvent exhibit
resistances in the M$\Omega$ range in the absence of any SAM,
indicating the formation of an insulating surface layer.
\end{thebibliography}
\end{document}